2018

# Systematic Mapping Protocol

COVERAGE OF ASPECT-ORIENTED METHODOLOGIES FOR THE EARLY PHASES OF THE SOFTWARE DEVELOPMENT LIFE CYCLE

FINAL VERSION: 2018/09/24

Fernando Piniroli[a], José Luis Barros Justo[b,] Laura Zeligueta[a], Marcelo Palma[a],

[a] Faculty of Informatics and Design, Champagnat University, Mendoza, Argentina.
[b] School of Informatics (ESEI), University of Vigo, Orense, Spain.

# Contents

# 1. Introduction

The amount of aspect-oriented software development techniques and tools have been increasing for the last years [1] [2] but still, they have not enough maturity and are not sufficiently spread to be included in a project leader's box of tools [3] [4]. Software development projects have to deal with many risks and, the main function of project leaders is to minimize the damage that these risks can cause. The use of immature technologies, tools newcomers to the market, techniques that have not been tested enough, etc., would be very risky decisions to take by who has the responsibility to carry out a successful software development project. On the other hand, the availability of well-known tools and techniques and the adherence to standards and best practices will help professionals to make good estimates and to take better decisions.

There are many proposals on aspect-oriented techniques, notations, tools, etc., but they have not yet been unified on a common body of knowledge and none of them has become the most popular approach.

We are interested in the early phases of the aspect-oriented software development life cycle which include the phases from the beginning of the life cycle until architecture design [5]. The phases considered in our research work will include: business model, request model, and requirement model; considering three views for the requirement model: functional, static and states view. We are interested in portraying the state of the art of aspect-oriented techniques and tools, in the identification of the standards they employ. Our goal is to collect all the available evidence, analyze it, and study the possibility of applying these techniques, tools and standards in real projects, taking advantage of the benefits of the aspect-oriented paradigm.

Evidence-Based Software Engineering (EBSE) aims to convert the need for information into an answerable question, tracking down the best evidence with which to answer that question and critically appraising the evidence for its validity [6]. Kitchenham et al. affirm that EBSE intends "to provide the means by which current best evidence from research can be integrated with practical experience and human values in the decision making process regarding the development and maintenance of software" [6]. In this document we detail the planning phase of a Systematic Mapping Study (SMS), used to structure the findings on a research area, based on the guidelines from Petersen et al. [7].

Our goal is twofold: to identify the standard and widespread approaches, techniques, notations and tools reported in the scientific literature, and to verify if they are applied in the industry.

We will perform a systematic mapping study of the literature up to July 2018.

The rest of the article is structured as follows: in Section 2 we describe the research method and Section 3 presents the strategy to deal with validity threats.

## 2. Research method

### 2.1. Goal and research questions

This work aims to identify and classify the aspect-oriented software development methodologies used to reduce the impact of moving from traditional approaches to the aspect-oriented approach.

It is of our particular interest to identify all the methodologies that could collaborate with this objective, regardless of whether they were already applied in the industry or if they are still in a study stage, without having reached their employment in real-settings.

Particularly we will consider:

- the early phases of the Software Development Life Cycle (SDLC) they encompass
- the notations and tools they use
- the approaches they propose
- impact they had on projects in real-world settings

A set of Research questions (RQ) has been designed to accomplish this goal (see Table 1). Furthermore, a set of publication questions (PQ) has been included to characterize the bibliographic and demographic space (

Table **2**).

Table 1. Research questions

| RQ# | Research question | Description |
|---|---|---|
| RQ1 | Which aspect-oriented methodologies (AOM) have been proposed? | A list of proposed aspect-oriented methodologies, for example: AORE, Theme and Approach, among others. |
| RQ2 | Which early phases of the SDLC have been covered by the AOM? | A list of the found early phases of the SDLC covered by the AOM, for example: business modeling, user requirement modeling, and software requirement modeling; considering three view for the last phase: functional, static and states. |
| RQ3 | Which notations are used by the AOM? | A list of notations used by AOM: UML, BPMN, etc. |
| RQ4 | Which modeling techniques are used by the AOM? | A list of modeling techniques used by AOM, for example: use cases, class or states models, etc. |
| RQ5 | What tools support the AOM? | Commercially available support tools, for example: Enterprise Architect, Visual Paradigm, etc. |

| RQ# | Research question | Description |
|---|---|---|
| RQ6 | Which of the identified AOM has been used in real-world settings (industry)? | A list of the AOM actually used by industry. |
| RQ7 | What are the benefits from the use of AOM on industry? | Identify reported benefits, for example: reusability, productivity, quality, cost reduction, understandability, ease of maintenance, etc. |
| RQ8 | What are the challenges of using AOM in real-world settings? | Identify potential research opportunities. |

Table 2. Publication questions

| PQ# | Publication question | Description |
|---|---|---|
| PQ1 | Where the studies had been published? | To know the distribution of studies by type of venue: conferences, journals or workshops. |
| PQ2 | How the quantity of studies has evolved? | Publications per year. |
| PQ3 | What are the authors' affiliations? | Classify the affiliations into two categories: academy or industry. We will consider the affiliations of all the authors. |
| PQ4 | Which are the most active countries? | Considering the author's affiliations (all authors). |

## 2.2. Search strategy and study selection

The selected search strategy includes two approaches to look for the primary studies. The first one is a manual search on the most important conference on aspect-oriented software development. The second one is an automatic search conducted through the online sources for scientific studies (digital libraries and databases). Finally, we will try to ensure the completeness of our set of studies by using the backward and forward snowballing technique [8]. Figure 1 shows these strategies.

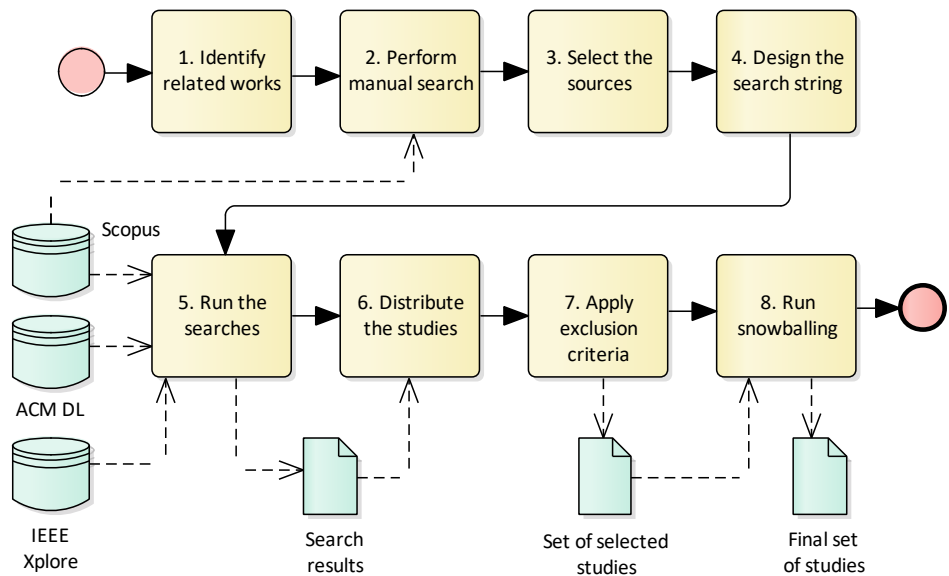

Figure 1. Search and selection process

The study selection strategy will include the classification and revision of every study in the set of retrieved works, aiming to select those relevant papers, regarding to the RQs, as presented in Figure 1 above.

The activities are as follows:

**Activity #1: Identify related works**
We will start our study identifying the related works previously conducted, as recommended by Petersen et al. [13], since this will also help us to adjust the focus of our study. As we are conducting a SMS, a type of secondary study, we will only consider as related work other secondary works (SMS or SLR) previously published.

**Activity #2: Perform manual search**
We have planned to conduct a manual search on Scopus, because it records all he peer-reviewed studies, on the major international conference related to the aspect-oriented approach: "AOSD – International Conference on Aspect-Oriented Software Development".

**Activity #3: Select the sources**
The electronic databases of scientific articles selected for this study are Scopus, IEEE Xplore and ACM digital library, as they are cited repeatedly in SMS reports and guidelines [9] [10] [11] [12].

**Activity #4: Design the search string for each source**
The search terms chosen that will be run mainly on title, abstract and keywords belong to the categories stated by the PICOC method (Petersen et al. [7]) as follows:

**Population:** we want to find tools, method, techniques, etc., that's why the selected terms are "(method* OR model* OR notation OR tool OR technique* OR UML OR BPMN)".

**Intervention:** this category refers to software engineering areas, so we have selected the early phases of the SDLC as follows: "(business model* OR analysis OR design OR architectur* OR requirement* OR early aspect*)". These terms include "early aspects" because it is the common way to denominate the use of the aspect-oriented approach in the early phases of the SDLC in the aspect-oriented community.

**Comparison:** we don't want to compare, since we just look for to list the identified results. Then, we don't have chosen terms for this category.

**Outcome:** we don't want to restrict the outcomes, so we don't have settled terms for this category.

**Context:** we have settled the aspect-oriented approach, so we include the terms "(aspect-orient* OR AOSD)".

The selected time frame starts on January 1996, because the aspect-oriented paradigm was born between November 1995 and May 1996. The time frame ends on July 2018.

The specific search strings for each database are the following ones:

Table 3. Search strings

| Database | Search string |
|---|---|
| Scopus | ( TITLE-ABS-KEY (method* OR model* OR notation OR tool OR technique* OR UML OR BPMN) AND TITLE-ABS-KEY ("business model" OR analysis OR design OR architectur* OR requirement* OR "early aspect") AND (aspect-orient* OR AOSD) ) AND PUBYEAR > 1995 AND PUBYEAR < 2019 AND ( LIMIT-TO(SUBJAREA,"COMP " ) ) AND ( LIMIT-TO(DOCTYPE,"cp " ) OR LIMIT-TO(DOCTYPE,"ar " ) ) AND ( LIMIT-TO(LANGUAGE,"English" ) ) |
| IEEE Xplore | ("Index Terms":method* OR "Index Terms":model* OR "Index Terms":notation* OR "Index Terms":tool* OR "Index Terms":technique* OR "Index Terms":UML OR "Index Terms":BPMN) AND ("Index Terms":"business model" OR "Index Terms":analysis OR "Index Terms":design OR "Index Terms":architectur* OR "Index Terms":requirement* OR "Index Terms":"early aspect") AND ("Index Terms":aspect-orient* OR "Index Terms":AOSD) |
| ACM | (acmdlTitle:(method* OR model* OR notation OR tool OR technique* OR UML OR BPMN) AND acmdlTitle:("business model" OR analysis OR design OR architectur* OR requirement* OR "early aspect") AND acmdlTitle:(aspect-orient* OR AOSD)) OR (recordAbstract:(method* OR model* OR notation OR tool OR technique* OR UML OR BPMN) AND recordAbstract:("business |

| Database | Search string |
|---|---|
| | model" OR analysis OR design OR architectur* OR requirement* OR "early aspect") AND recordAbstract:(aspect-orient* OR AOSD)) OR (keywords.author.keyword:(method* OR model* OR notation OR tool OR technique* OR UML OR BPMN) AND keywords.author.keyword:("business model" OR analysis OR design OR architectur* OR requirement* OR "early aspect") AND keywords.author.keyword:(aspect-orient* OR AOSD)) |

**Activity #5: Run the searches**
The searches are executed and the results collected. These results will contain duplicates that must be eliminated by applying these rules:
   a. Expanded works (or expanded versions): keep the last one.
   b. Duplicated works: depending on the source, following this priority order: Scopus (since it offers the most detailed information), followed by IEEE Xplore and, finally, ACM DL (because it does not retrieve the abstracts of the studies) [14].

**Activity #6: Distribute the studies**
The retrieved studies will be distributed among four researchers as Table 4 shows. Notice that we ensure that every single work will be examined by two different researchers, in order to reduce bias.

Table 4. Distribution of studies.

| Researcher | Studies | | | |
|---|---|---|---|---|
| | 0%-25% | 26%-50% | 51%-75% | 76%-100% |
| R1 | X | X | | |
| R2 | | X | X | |
| R3 | | | X | X |
| R4 | X | | | X |

The individual selection of studies made by each researcher will be consolidated into a unique set of studies. Differences among researchers will be solved by using the following criteria [7]:

Table 5. Criteria to resolve disagreements.

| | | Researcher 1 | | |
|---|---|---|---|---|
| | | Include | Uncertain | Exclude |
| Researcher 2 | Include | A | B | D |
| | Uncertain | B | C | E |

|  | Exclude | D | E | F |
|---|---|---|---|---|

**A & B:** the study is included.
**E & F:** the study is excluded.
**C & D:** the paper is read in full and qualified again until obtaining A, B, E or F.

**Activity #7: Apply exclusion criteria**
The researchers will independently review the studies they have been assigned to and they will decide whether the studies are relevant or not, by only reading their title and abstract and then applying the exclusion criteria (EC). The set of retrieved studies will be then filtered by applying the exclusion criteria described below, in Table 6.

Table 6. Exclusion criteria.

| EC# | Description |
|---|---|
| EC1 | The study is not written in English. |
| EC2 | The study venue is not conference, workshop or journal. |
| EC3 | The study is not peer-reviewed. |
| EC4 | Short papers (less than four pages). |
| EC4 | The focus is not on an aspect-oriented methodology for the early SDLC phases. |
| EC5 | The focus is not on an aspect-oriented technique, tool or notation. |

**Activity #8: Run snowballing**
Resulting articles will be considered as "seed works" to be used on a forward and a backward snowballing technique, following the guidelines proposed by Wohlin [8]. The motivation for running a forward and backward snowballing complementary search aims to complement the automatic search and to collaborate with the search strings refinement.

## 2.3. Data extraction form

Relevant data are extracted from the set of selected studies to answer the eight RQs and the four PQs. Data are stored into a spreadsheet with the format shown in Table 7 (Data Extraction Form, DEF) and in Table 8.

Table 7. Data extraction form for RQ.

| Study #ID | RQ1 | RQ2 | RQ3 | RQ4 | RQ5 | RQ6 | RQ7 | RQ8 |
|---|---|---|---|---|---|---|---|---|
| Study #1 | | | | | | | | |
| Study #2 | | | | | | | | |
| … | … | … | … | … | … | … | … | |
| Study #n | | | | | | | | |
| **Accepted values** | Methodology names (text) | Phase names (text) | Notation names (text) | Technologies names (text) | Tool names (text) | Methodology names (text) | Benefit names (text) | Challenge names (text) |
| **Chart type** | Bar | Pie | Pie | Pie | Bar | Bar | Bar | Bar |

We have selected different presentations depending on the amount of possible results: when they may be a lot, we will use a bar chart, but we will employ a pie chart when could be a few.

Table 8. Data extraction form for PQ.

| Study #ID | PQ1 | PQ2 | PQ3 | PQ4 |
|---|---|---|---|---|
| Study #1 | | | | |
| Study #2 | | | | |
| … | … | … | … | … |
| Study #n | | | | |
| **Accepted values** | Fora names (text) | Year of publication (integer) | Academia Industry Research center | Country names (text) |
| **Chart type** | Bar | Line | Pie | Pie |

# 3. Threats to validity

In order to minimize the impact of the validity threats categorized by Petersen [7] that could affect our study, we present them with the corresponding mitigation actions:

**Descriptive validity**

This validity seeks to ensure that observations are objectively and accurately described.

- We have structured the information to be collected by means of a couple of Data Extraction Forms, for RQs and PQs, presented in Table 7 and Table 8, to support an uniform recording of data and to objectify the data extraction process.

- Besides, all the researchers will participate on an initial meeting, aimed at unifying concepts and criteria, answer to any question and to demonstrate (by examples) how to conduct the data extraction process.
- We will also make public our data extraction form.

**Theoretical validity**

The theoretical validity depends on the ability to get the information that it is intended to capture.

- We will start with a search string (Table 3) tailored for the three most popular digital libraries on computer sciences and software engineering online databases.
- An expert will provide a set of articles to verify if they are retrieved with the search string.
- A set of exclusion criteria (Table 6) to objectivize the selection process have been defined.
- We will distribute the studies among four researchers, working independently and, with an overlap of studies that ensures that each study is reviewed by at least two researchers (Table 4).
- We will combine two different search methods: an automatic search and a manual search (backward and forward snowballing), to diminish the risk of not finding all the available evidence.
- It could be a minimal impact due to the selection of articles written in English and the discard of other languages.

**Generalizability**

This validity is concerned with the ability to generalize the results to the whole domain.

- Our set of RQs is general enough in order to identify and classify the findings on aspect-oriented software development methodologies regardless specific cases, type of industry, etc. [15]

**Interpretive validity**

This validity is achieved when the conclusions are reasonable given the data.

- At least two researchers will validate every conclusion.
- Two researchers, experienced on the problem domain, will help us with the interpretation of data.

**Repeatability**

The research process must be detailed enough in order to ensure it can be exhaustively repeated.

- We have designed this protocol sufficiently detailed to allow to repeat the process we have followed.
- The protocol, as well as the results of the study, will be published online, so other researchers can replicate the process and, hopefully, corroborate the results.

## 4. Conclusions

We have strictly followed the guidelines published by Petersen [7] to: plan, conduct and report a SMS. As the whole team adhered to these guidelines to build up the protocol presented in this document, we think that the execution phase (conducting the SMS) will be repeatable and that the threats to validity have been mitigated as much as possible.